\documentclass[aps, prd, twocolumn, superscriptaddress, preprintnumbers, nofootinbib, longbibliography]{revtex4-1}

\usepackage{amsmath}	
\usepackage{amsthm}		
\usepackage{amssymb}	
\usepackage{datetime}
\usepackage{graphicx}
\usepackage{color}
\usepackage{verbatim}
\usepackage{bm}

\usepackage[colorlinks=true, citecolor=midblue, linkcolor=midblue, urlcolor=midblue]{hyperref}

\usepackage{setspace}

\usepackage[svgnames]{xcolor}

\usepackage{tikz,xcolor}
\definecolor{lime}{HTML}{A6CE39}
\DeclareRobustCommand{\orcidicon}{
	\begin{tikzpicture}
	\draw[lime, fill=lime] (0,0) 
	circle [radius=0.16] 
	node[white] {{\fontfamily{qag}\selectfont \tiny ID}};
	\draw[white, fill=white] (-0.0625,0.095) 
	circle [radius=0.007];
	\end{tikzpicture}
	\hspace{-2mm}
}
\foreach \x in {A, ..., Z}{
	\expandafter\xdef\csname orcid\x\endcsname{\noexpand\href{https://orcid.org/\csname orcidauthor\x\endcsname}{\noexpand\orcidicon}}
}

\settimeformat{ampmtime}

\definecolor{grey}{rgb}{0.4,0.4,0.4}
\definecolor{dullmagenta}{rgb}{0.4,0,0.4}
\definecolor{darkblue}{rgb}{0,0,0.4}
\definecolor{midblue}{rgb}{0,0,0.5}
\definecolor{midred}{rgb}{0.5,0,0}
\definecolor{orange}{rgb}{1,0.5,0}
\definecolor{lightbrown}{rgb}{0.75,0.5,0.25}
\definecolor{tan}{cmyk}{0.14,0.42,0.56,0}
\definecolor{djunglegreen}{cmyk}{0.99,0,0.52,0}
\definecolor{lightgreen}{rgb}{0,1,0}
\definecolor{olivegreen}{cmyk}{0.64,0,0.95,0.40}
\definecolor{midgreen}{rgb}{0.0,0.675,0.0}
\definecolor{darkgreen}{rgb}{0,0.5,0}

\settimeformat{ampmtime}

\usepackage[font=small,labelfont=bf,justification=raggedright]{caption}
\usepackage{subcaption}

\usepackage[notrig]{physics}
\usepackage{tensor}

 \newcommand{\ThirdAffiliation}{\affiliation{
        INFN, Sezione di Pisa, Largo Bruno Pontecorvo 3, I-56127 Pisa, Italy
	}}

\date{\formatdate{\day}{\month}{\year}, \currenttime}

\begin{document}

\title{On the predictivity of axion dark matter in the presence of Peccei–Quinn breaking}

\author{Michael Zantedeschi\orcidC{}}
\email{michael.zantedeschi@pi.infn.it}
\ThirdAffiliation

\date{\small\today} 

\begin{abstract}
It is shown that the post-inflationary quantum chromodynamics (QCD) axion
need not lead to a unique one-parameter prediction for the dark matter
abundance whenever small explicit Peccei--Quinn symmetry breaking becomes
dynamically relevant before the QCD transition.
Although strongly constrained by the strong CP bound, such breaking remains phenomenologically viable and introduces a mass scale $\mu$ that can control the early-time dynamics, as the QCD contribution to the axion mass is thermally suppressed at high temperatures. In this case, the axion string--wall network annihilates earlier, and the relic abundance is no longer primarily set by QCD dynamics alone, but instead depends on $\mu$, in addition to $f_a$, the axion decay constant. This effect overlaps with the parameter space relevant for QCD axion dark matter and, depending on ultraviolet parameters and initial conditions, can extend across it entirely.
\end{abstract}

\maketitle

\section{Introduction}

QCD is characterized by a continuum of vacua~\cite{Callan:1976je,Jackiw:1976pf}
labeled by the vacuum angle $\vartheta$. This CP-violating quantity
denotes different superselection sectors. However, the physical parameter
is given by $\overline{\vartheta} = \vartheta + \arg\det M$, where
$\arg\det M$ is the phase of the quark mass matrix determinant.

At the quantum level, $\overline{\vartheta}$ induces an electric dipole
moment of the neutron~\cite{Baluni:1978rf,Crewther:1979pi}, whose current
experimental limit, $|d_n| < 2.9 \times 10^{-26}\,\mathrm{e\,cm}$~\cite{Baker:2006ts},
implies
\begin{equation}
\label{eq:varthetabound}
    \overline{\vartheta} \lesssim 10^{-10}\,.
\end{equation}
Additional contributions to the neutron electric dipole moment arise from
CP violation in the electroweak sector~\cite{Ellis:1976fn,Shabalin:1979gh,Ellis:1978hq},
but these are negligibly small and therefore irrelevant for the present
discussion. In this sense, $\overline{\vartheta} \ll 1$ can be regarded as
an input parameter of the Standard Model~\cite{Senjanovic:2020pnq}.
Nevertheless, its smallness has long been considered puzzling.

Remarkably, $\overline{\vartheta}$ can be promoted to a dynamical field,
the axion, which relaxes the vacuum to a CP-invariant ground state. This
new particle has been extensively studied as a dark matter candidate and
is actively searched for.

In its original formulation, by Peccei and Quinn (PQ)~\cite{Weinberg:1977ma,Wilczek:1977pj,Peccei:1977hh},
the axion is a pseudo-Nambu--Goldstone boson of an anomalous $U(1)_{\rm PQ}$
symmetry. This mechanism dynamically relaxes $\overline{\vartheta}$ to zero.
However, it replaces an extremely small parameter with a highly accurate
global symmetry, which must be broken only by the QCD anomaly. Since
global symmetries are not expected to be exact, this leads to the axion
quality problem.

From an effective field theory perspective, additional PQ-violating effects are generically expected from ultraviolet physics. Related non-QCD contributions to the axion potential, and their cosmological implications, have also been explored in alternative frameworks, such as non-compact axion scenarios~\cite{Karananas:2025uhy}. Requiring compatibility with Eq.~\eqref{eq:varthetabound} implies that such effects must be parametrically suppressed relative to the QCD contribution. 

Cosmological implications of explicit PQ breaking, including its impact on topological defects and axion dynamics, have also been studied in various contexts, for instance in aligned axion models, see e.g.,~\cite{Higaki:2015jag,Higaki:2016yqk}.

Alternative formulations of the axion without global symmetries have been
proposed~\cite{Dvali:2005an,Dvali:2022fdv}. Here we focus on the standard
PQ framework, which is the primary target of current experimental
searches~\cite{DiLuzio:2020wdo}.

The cosmological prediction of the axion relic abundance depends on the
thermal history of PQ symmetry breaking. Axions can be produced
non-thermally via the misalignment mechanism
\cite{Preskill:1982cy,Abbott:1982af,Dine:1982ah}.
In the pre-inflationary scenario, the axion field is homogenized by
inflation and the relic abundance depends on the initial misalignment
angle, which is not predicted by the theory. This scenario is therefore
intrinsically non-predictive.

In the post-inflationary scenario, the axion field takes random values in
causally disconnected regions, leading to the formation of a cosmic
string network via the Kibble mechanism
\cite{Kibble:1976sj,Vilenkin:1982ks,Lyth:1992tw}. As the Universe
expands, the network evolves and radiates energy predominantly into
axions. Around the QCD epoch, the axion potential turns on, domain walls
form, and the string--wall system eventually annihilates, producing cold
axions
\cite{Chang:1998tb,Hiramatsu:2010yn,Hiramatsu:2012gg,Kawasaki:2018bzv,Saikawa:2012uk,Vaquero:2018tib}.

In this case, the relic abundance is often regarded, up to the uncertainties associated with the
string network, as a function of the single parameter $f_a$, leading to an approximate one-to-one mapping
between $f_a$ and the dark matter density
\cite{Gorghetto:2018myk,Gorghetto:2020qws,Hindmarsh:2021zkt,Hiramatsu:2012sc,Buschmann:2021sdq,Benabou:2024msj}.

In this work we show that this predictivity is generically lost once explicit Peccei--Quinn breaking becomes dynamically relevant before the QCD transition. Even if
compatible with the strong CP bound, such effects introduce an additional
timescale which can determine the annihilation time of the string
network. When this occurs before the QCD transition, the cosmological
evolution is no longer controlled primarily by infrared QCD dynamics.

The main message of this work is the following:
\textit{The axion relic abundance need not be an infrared prediction of
QCD, but instead probes ultraviolet Peccei--Quinn symmetry breaking once
explicit violations become dynamically relevant before the QCD transition.
In this regime, the one-to-one mapping between the axion mass and the dark matter
abundance is lost.}

In the next section we recall how explicit Peccei--Quinn breaking beyond
the QCD anomaly reintroduces a physical $\overline{\vartheta}$. Section~III
presents the cosmological estimates, and Section~IV contains the
conclusions.

\section{Topological susceptibility, pole structure, and explicit breaking}

In the following we shall follow the notation of \cite{Dvali:2022fdv}. The existence of $\vartheta$-vacua in QCD is equivalent to a non-vanishing
topological susceptibility (TSV), reflecting the sensitivity of the vacuum energy to the CP-violating angle. This can be formulated in terms of the
topological density
\begin{equation}
E(x)\equiv \frac{g_s^2}{32\pi^2} F_{\mu\nu}^a \tilde F^{a\,\mu\nu},
\end{equation}
whose spacetime integral measures the topological charge. The TSV is the
zero-momentum limit of the $E$--$E$ correlator,
\begin{equation}
\chi \equiv \int d^4x\,\langle E(x)E(0)\rangle
\;=\;\lim_{p\to 0}\,\langle E\,E\rangle(p),
\end{equation}
which is of order $\Lambda_{\rm QCD}^4$, with $\Lambda_{\rm QCD}\sim 200\,\mathrm{MeV}$~\cite{Witten:1979vv,Veneziano:1979ec}. Physically, $\chi$ measures the curvature of the vacuum energy as a function of $\vartheta$.

A key point is that $E$ is a total derivative. Introducing the Chern--Simons
three-form $C$ (or equivalently the Chern--Simons current), one has
\begin{equation}
E = {\rm d}C \qquad \text{(equivalently, } E = \partial_\mu C^\mu \text{)}.
\end{equation}
In momentum space this implies 
$\langle E\,E\rangle(p) \;=\; p^2\,\langle C\,C\rangle(p)$
up to tensor structures that are irrelevant for the infrared discussion.
Therefore, a non-vanishing susceptibility $\chi=\lim_{p\to 0}\langle E E\rangle(p)\neq 0$
is equivalent to the statement that $\langle C\,C\rangle$ contains a massless pole. This is the field-theoretic manifestation of the continuum of $\vartheta$-vacua.

In particular, the K\"all\'en--Lehmann representation for the Chern--Simons
three-form correlator takes the form
\begin{equation}
\langle C\,C\rangle(p)
=
\frac{\rho(0)}{p^2}
+
\sum_{m\neq 0}\frac{\rho(m^2)}{p^2-m^2},
\qquad \rho(0)\neq 0,
\end{equation}
where the massive modes are separated from the $p^2=0$ pole by a finite gap.
The pole at $p^2=0$ encodes the continuum of $\vartheta$-vacua, while the massive modes are irrelevant in the infrared. Correspondingly, $\langle E E\rangle(p)$ approaches a constant as $p\to 0$,
in agreement with $\chi\neq 0$.

This pole structure implies that the infrared effective action must reproduce this behavior. Since the massless pole resides in the correlator of the Chern--Simons three-form, the infrared EFT must contain a kinetic term for this three-form. When written
in terms of the gauge-invariant field strength $E={\rm d}C$, this is captured by~\cite{Dvali:2022fdv}
\begin{equation}
\mathcal L_{\rm eff}\supset \mathcal{K}(E),
\qquad
\mathcal{K}(E)=\frac{1}{2\chi} E^2+\cdots,
\end{equation}
so that $\chi$ sets the normalization of the quadratic term in the effective action.

The axion couples to the topological density through
\begin{equation}
\mathcal L_{\rm int} = - \frac{a}{f_a} E,
\end{equation}
so that the low-energy effective theory takes the form
\begin{equation}
\mathcal L = \mathcal{K}(E) + \frac{1}{2} (\partial_\mu a)^2 - \frac{a}{f_a} E.
\end{equation}
The function $\mathcal{K}(E)$ is such that, in the absence of explicit
PQ breaking, the effective theory reproduces the expected
periodic dependence on the axion field. In the following we work in a
local expansion around a given branch and retain only the leading
quadratic term, $\mathcal{K}(E)=E^2/(2\chi)$, which is sufficient
for our purposes. This corresponds to expanding the periodic potential
around a given $\vartheta$-branch; global properties such as periodicity
and domain wall number are not resolved in this approximation.

The three-form field does not carry propagating degrees of freedom in four
dimensions. As a result, its field strength $E={\rm d}C$ can be integrated out through
its equation of motion. At quadratic order one finds
\begin{equation}
E = \chi \frac{a - a_0}{f_a},
\end{equation}
where $a_0$ is an integration constant labeling the different $\vartheta$-branches. 
In the absence of the axion, this constant corresponds to the physical CP-violating angle,
\begin{equation}
\bar\vartheta_v \equiv a_0/f_a\,.
\end{equation}
Substituting back into the Lagrangian yields the effective potential
\begin{equation}
V(a) = \frac{\chi}{2 f_a^2} (a - a_0)^2 + \cdots.
\end{equation}
Therefore, the axion acquires a mass $m_a^2 = \chi/f_a^2$, with the minimum at $a=a_0$ (correspondingly, $E=0$ is screened).

The important consequence is that the coupling to the axion gaps the
massless pole in the Chern--Simons three-form correlator. The axion plays
the role of a St\"uckelberg field, and the three-form becomes massive, with a gap set by $m_a$~\cite{Dvali:2005an}. As a result, the correlator takes the form
\begin{equation}
\langle C C \rangle(p) \sim \frac{1}{p^2 - m_a^2},
\end{equation}
so that the pole at $p^2=0$ is lifted.
Since $E = {\rm d}C$, one has
$
\langle E E \rangle(p) \sim p^2 \langle C C \rangle(p)
\;\sim\; {p^2}/(p^2 - m_a^2),
$
which vanishes in the infrared limit. 
Consequently, the physical topological susceptibility is screened,
$\lim_{p \to 0} \langle E E \rangle(p) = 0$, 
corresponding to the dynamical relaxation of the effective $\bar \vartheta$-angle to zero.

Let us now introduce a small explicit breaking of the axion shift symmetry.
For the purpose of deriving the induced residual CP violation, it is
sufficient to work locally around one branch, where the leading deformation
can be parametrized as
\begin{equation}
\Delta \mathcal L = - \frac{1}{2} \mu^2 a^2 .
\end{equation}
In a complete ultraviolet theory, the explicit breaking is generically
periodic and characterized by a domain-wall number, but this global
structure is not relevant for the present local analysis.

Integrating out $E$ as before yields
\begin{equation}
V(a) = \frac{\chi}{2 f_a^2} (a - a_0)^2 + \frac{1}{2} \mu^2 a^2.
\end{equation}
Minimizing the potential gives
\begin{equation}
a_{\rm min} = \frac{a_0}{1 + \mu^2 f_a^2/\chi},
\end{equation}
and therefore the vacuum expectation value of the topological density is
\begin{equation}
E_0 \equiv E(a_{\rm min})
=
- \chi\,\frac{a_0}{f_a}\,
\frac{\mu^2 f_a^2}{\chi+\mu^2 f_a^2}.
\end{equation}
Thus, for $\mu \neq 0$, the theory retains a residual sensitivity to the
initial CP-violating angle $\bar\vartheta_v$, and a non-vanishing vacuum expectation value is induced. In the absence of explicit breaking $\bar\vartheta_v$ would be unphysical, but once the shift symmetry is broken it becomes a physical parameter controlling the residual CP violation.

In the regime $\mu^2 \ll m_a^2$, one finds
\begin{equation}
E_0 \simeq - \bar\vartheta_v \, \mu^2 f_a^2 .
\end{equation}
This corresponds to an induced effective CP-violating angle
\begin{equation}
\bar\vartheta \equiv \bar\vartheta_v \, \frac{\mu^2 f_a^2}{\chi}.
\end{equation}
Correspondingly, explicit breaking prevents complete screening of the original CP-violating angle.

The neutron electric dipole moment bound \eqref{eq:varthetabound} therefore implies
\begin{equation}
\bar\vartheta_v \, \frac{\mu^2 f_a^2}{\chi} \lesssim 10^{-10}.
\end{equation}
In particular, even if $\bar\vartheta_v \sim \mathcal O(1)$, the strong CP problem is still solved provided the explicit breaking is sufficiently small.
Equivalently, in terms of the axion mass, this condition can be written as
\begin{equation}
\label{eq:muma1010}
\frac{\mu^2}{m_{a}^2}
=
\frac{\bar \vartheta}{\bar\vartheta_v}
\lesssim
\frac{10^{-10}}{\bar\vartheta_v}.
\end{equation}
Since $\bar\vartheta_v$ is not observable in the absence of explicit breaking, its value is \textit{a priori} unknown. Even if extremely small,
it introduces a new physical timescale which can dominate the cosmological
evolution of the axion field.

\section{Cosmological implications}

We now discuss the cosmological implications of a small explicit breaking
of the axion shift symmetry. The key observation is that, even if $\mu$
is sufficiently small to satisfy the strong CP bound, it introduces a new
physical timescale $\mu^{-1}$ which can control the early-time evolution
of the axion field. This is possible because the QCD contribution to the
axion potential is strongly thermally suppressed at high temperatures,
whereas the explicit breaking considered here is approximately
temperature-independent. As a result, even a very small explicit breaking
can dominate the dynamics before the QCD transition. Similar effects, in which subleading PQ-breaking contributions control
the axion evolution at early times, have been discussed in the literature,
primarily in the pre-inflationary scenario~\cite{Jeong:2022kdr}, and in
frameworks with additional non-QCD contributions to the axion potential~\cite{Karananas:2025uhy}.

We focus on the post-inflationary PQ scenario, in which the axion field
takes random values in causally disconnected regions and a network of
global strings forms when the PQ symmetry is spontaneously broken.

In the absence of explicit breaking, the subsequent evolution is controlled
by QCD. Around the confinement scale, the QCD-induced axion potential
turns on, domain walls form, and the string--wall network annihilates at a
temperature $T \simeq \Lambda_{\rm QCD}$. In this case the resulting axion
abundance is often regarded, up to the uncertainties associated with the
string network, as a function of the single parameter $f_a$. The
temperature dependence of the QCD axion potential and its implications for
cosmology have been analysed in detail using chiral effective field theory
and lattice inputs; see \cite{diCortona:2016AxionPrecisely} and the updated
precision normalisation of $m_a$ in \cite{Gorghetto:2019TopSus}.

For the cosmological discussion the global structure of the potential is
important. We therefore write schematically
\begin{equation}
\label{eq:fullpot}
V(a,T)=
\chi(T)\left[1-\cos\!\left(\frac{a}{f_a}\right)\right]
+
V_{\rm br}(a),
\end{equation}
where the QCD topological susceptibility $\chi(T)$ is strongly suppressed
at high temperature. A generic explicit-breaking contribution can be
parametrized as
\begin{equation}
V_{\rm br}(a)
=
\mu^2 f_a^2
\left[
1-\cos\!\left(N_{\rm br}\frac{a}{f_a}+\delta\right)
\right],
\end{equation}
where $N_{\rm br}$ and the phase $\delta$ depend on the ultraviolet
operator responsible for the breaking. We take the QCD anomaly to
correspond to the standard post-inflationary case with
$N_{\rm DW}^{\rm QCD}=1$, while $N_{\rm br}$ parametrizes the periodicity
of the explicit-breaking term relative to the QCD potential. The quadratic
term used in Sec.~II corresponds to a local expansion of this potential
around a given branch. By contrast, the global quantities $N_{\rm br}$ and
$\delta$ determine the relative alignment of the two contributions and
whether the explicit breaking lifts or preserves degeneracies among the
vacua.

The explicit breaking lifts the degeneracy among the vacua and induces an
energy bias of order
\begin{equation}
\Delta V \simeq \mu^2 f_a^2 ,
\end{equation}
up to model-dependent numerical coefficients. This bias provides a pressure
difference across domain walls separating inequivalent vacua and can drive
the collapse of the string--wall network.
 At the parametric level, this effect can be
characterized by an effective scale governing the force on the
string--wall network,
\begin{equation}
\sigma_{\rm exp} \simeq \mu f_a^2 ,
\end{equation}
again up to order-one factors. If the explicit breaking is characterized
by a domain wall number $N_{\rm br}$, the corresponding numerical
coefficients may depend on $N_{\rm br}$, but this does not modify the
parametric estimates below.

In what follows we assume that the explicit breaking selects a unique
vacuum for the relevant post-inflationary network, so that no stable
domain-wall network remains. If the relative phase $\delta$ and the
integer $N_{\rm br}$ lead instead to metastable configurations, the
annihilation time and the emitted spectrum can be modified. A complete
treatment of such cases requires dedicated numerical simulations of the
full periodic potential.

These walls pull on the string network, whose tension is
\begin{equation}
\mu_{\rm str} \simeq \pi f_a^2 \log\!\left(\frac{m_\rho}{H}\right),
\end{equation}
where $m_\rho \simeq f_a$ is the mass scale of the radial mode and the
infrared cutoff is set by the Hubble scale. For a string with curvature
radius of order $H^{-1}$, the restoring force per unit length is of order
$\mu_{\rm str} H$, while the force exerted by the wall is of order
$\sigma_{\rm exp}$. The network therefore annihilates when
\begin{equation}
H_{\rm ann}
\simeq
\frac{\sigma_{\rm exp}}{\mu_{\rm str}}
\simeq
\frac{\mu}{\pi \log(m_{\rho}/H_{\rm ann})}\,,
\end{equation}
where $H_{\rm ann}$ is the Hubble parameter at the annihilation time.

This condition should be understood as a parametric estimate. It captures
the force balance between the wall tension pulling on the strings and the
restoring force associated with the logarithmically enhanced string
tension. A full network evolution can shift the annihilation time through
the detailed interplay of bias energy, wall tension, and the energy stored
in the scaling string network. We absorb these residual uncertainties into
the parameter $\kappa$ defined below.

The effect discussed here becomes relevant when the annihilation of the
string network occurs before the QCD transition, i.e.\ before the
QCD-induced axion potential becomes dynamically relevant. 
In this regime, the annihilation time of the network is controlled by explicit Peccei--Quinn symmetry breaking rather than QCD if
\begin{equation}
\label{eq:boundmu}
\mu \gtrsim \kappa\,H_{\rm QCD},
\end{equation}
where $\kappa$ encodes the logarithmic enhancement of the string tension
together with residual order-one numerical factors.
Parametrically,
\begin{equation}
\kappa \sim \pi \log\!\left(\frac{m_\rho}{H_{\rm ann}}\right)
\sim \mathcal O(10^2)\,.
\end{equation}
Numerically, for $f_a \sim 10^{10}$--$10^{12}\,\mathrm{GeV}$ and
$H_{\rm ann}\sim H_{\rm QCD}$, one finds
$\log(m_\rho/H_{\rm ann})\sim 60$--$70$, so that
$\kappa \sim 200$ up to order-one uncertainties. Here
$H_{\rm QCD}\simeq 10^{-19}\,\mathrm{GeV}$ is the Hubble parameter at
the QCD phase transition, corresponding to
$T_{\rm QCD}\sim 0.1$--$0.2\,\mathrm{GeV}$.

Combining Eq.~\eqref{eq:boundmu} with the strong CP constraint
in Eq.~\eqref{eq:muma1010}, one finds parametrically
\begin{equation}
\label{eq:res1}
m_a \gtrsim 10^{-3}\,
|\bar\vartheta_v|^{1/2}\,\mathrm{eV},
\end{equation}
up to logarithmic and order-one uncertainties. Using the standard
relation $m_a \propto f_a^{-1}$, this corresponds to
\begin{equation}
\label{eq:res2}
f_a \lesssim
10^{10}\,
|\bar\vartheta_v|^{-1/2}\,
\mathrm{GeV}.
\end{equation}
Therefore, for axion decay constants
$f_a \lesssim 10^{10}\,\mathrm{GeV}$ and
$|\bar\vartheta_v|\sim\mathcal O(1)$, explicit PQ breaking can trigger the
annihilation of the string network before the QCD transition. In this
regime the relic abundance is no longer controlled primarily by the
infrared QCD timescale, but instead by the explicit breaking scale.

For smaller values of $|\bar\vartheta_v|$, the bounds in
Eqs.~\eqref{eq:res1} and~\eqref{eq:res2} are correspondingly relaxed. In
the limit $\bar\vartheta_v \ll 1$, the strong CP bound places only a weak
restriction on $\mu$, and the region in which explicit breaking can affect
the cosmological evolution correspondingly broadens.

To support the above parametric result, we now estimate the relic axion
abundance in the regime where the decay of the network is driven by the
explicit breaking scale $\mu$. In the scaling regime, the energy density
stored in strings is
\begin{equation}
\label{eq:rhostr}
\rho_{\rm str} \simeq \xi\,\mu_{\rm str} H^2 \, ,
\end{equation}
where $\xi\simeq \mathcal{O}(1)$ parametrizes the number of long strings
per Hubble volume; see e.g.~\cite{Hindmarsh:2021zkt}. At annihilation,
this energy is converted into axions. The typical momentum of the
emitted axions is not fixed by the parametric argument alone and depends
on the detailed network evolution. We therefore write
\begin{equation}
\omega \sim q_\omega H_{\rm ann},
\end{equation}
where $q_\omega\gtrsim \mathcal O(1)$ parametrizes the characteristic
momentum in units of the Hubble scale. The estimate used below corresponds
to $q_\omega\sim\mathcal O(1)$; a harder spectrum, for instance with
$\omega\sim \mu$, would correspond to $q_\omega\sim \kappa$.
Therefore, the axion number density at annihilation is estimated as
\begin{equation}
n_a(t_{\rm ann})
\sim
\frac{\rho_{\rm str}(t_{\rm ann})}{\omega}
\sim
{\frac{\xi}{ q_\omega}}\,
\mu_{\rm str}\,H_{\rm ann}
\sim
{\frac{\xi}{q_\omega}}\,
\frac{\mu_{\rm str}}{\kappa}\,\mu .
\end{equation}
The ingredients entering this estimate -- the scaling properties of the
string network and the spectrum of emitted axions -- have been studied
extensively in lattice simulations; see e.g.
\cite{Gorghetto:2018Attractive,Gorghetto:2021MoreAxions}. Our purpose here
is not a precision prediction, but a parametric estimate of how the relic
abundance depends on the explicit breaking scale. The normalization
of this estimate, and in detailed models also the effective scaling, can
be modified by the emitted spectrum. The robust conclusion is that the
abundance depends on additional data beyond $f_a$ alone.

The corresponding yield is
\begin{equation}
Y_a \equiv \frac{n_a}{s}
\sim
{\frac{\xi}{q_\omega}}\,
\frac{\mu_{\rm str}}{M_{\rm Pl}^{3/2}}
\,\mu^{-1/2}\,\kappa^{1/2},
\end{equation}
where we used $s \sim g_* T^3$ and
$T_{\rm ann} \sim \sqrt{H_{\rm ann}M_{\rm Pl}}
\sim \sqrt{\mu M_{\rm Pl}/\kappa}$
during radiation domination.
The present-day abundance is then
\begin{equation}
\Omega_a h^2 \simeq \frac{s_0}{\rho_c/h^2}\, m_{a}\, Y_a,
\end{equation}
where $s_0$ is the present entropy density and $\rho_c$ is the critical
density. Using
$m_a \propto f_a^{-1}$ and $\mu_{\rm str} \simeq f_a^2$ up to logarithmic
corrections, one finds the parametric scaling
\begin{equation}
\Omega_a \propto \,f_a\,\mu^{-1/2}.
\end{equation}
Even in the standard post-inflationary scenario, the relic abundance is
affected by sizable uncertainties associated with the scaling regime, the
string density, and the spectrum of emitted axions
\cite{Gorghetto:2018myk,Gorghetto:2020qws,Hindmarsh:2021zkt,Hiramatsu:2012sc,Buschmann:2021sdq,Benabou:2024msj}.
Our use of the word predictivity should therefore be understood in this
limited sense: the minimal scenario gives an approximate one-parameter
relation between $f_a$ and the relic abundance. Explicit PQ breaking
introduces additional independent data, such as $\mu$, $N_{\rm br}$,
$\delta$, and the emitted spectrum, so that this one-parameter relation is
no longer fixed by infrared QCD dynamics alone.

The above estimate also neglects the subsequent evolution through the QCD
crossover. Axions produced before the QCD transition may initially be
relativistic, while both the axion mass and the position of the minimum
evolve as $\chi(T)$ turns on. If the explicit-breaking minimum is not
aligned with the QCD minimum, the motion of the minimum can induce
additional coherent oscillations and modify the final cold abundance. A
quantitative prediction therefore requires solving the real-time evolution
in the full temperature-dependent potential \eqref{eq:fullpot},
including the axion spectrum emitted by the network. Related effects in
time-dependent axion potentials and non-standard cosmological histories
have been studied in Refs.~\cite{Turner:1985si,Lyth:1991ub,Co:2018mho,Nelson:2018via,Visinelli:2017imh,Arias:2022qjt,Arias:2023wyg,Chang:2019tvx,Blinov:2019jqc}.
In this work we do not attempt a precision computation of the final
abundance; rather, we emphasize the parametric fact that the additional
scale $\mu$ can control the network annihilation time.

This has an important conceptual consequence: once explicit breaking
controls the annihilation time, the approximate one-parameter relation
between $f_a$ and the relic abundance is no longer fixed by infrared QCD
dynamics alone, but depends on ultraviolet data such as $\mu$, $N_{\rm br}$,
$\delta$, the emitted spectrum, and the thermal history.

As an illustrative benchmark, take the explicit breaking to saturate the
strong CP bound and assume $\bar\vartheta_v \sim \mathcal{O}(1)$, so that
$\mu \sim 10^{-5} m_a$. For a soft emission spectrum,
$q_\omega \sim \mathcal{O}(1)$, the parametric estimate above reproduces
the observed dark matter abundance for
$m_a \sim 10^{-3}\,\mathrm{eV}$, corresponding to
$f_a \sim 10^{10}\,\mathrm{GeV}$, in broad agreement with
Eqs.~\eqref{eq:res1} and~\eqref{eq:res2}. This numerical value should be
regarded as indicative, since it is sensitive to order-one factors, the
emitted axion spectrum, and the subsequent evolution through the QCD
crossover. For smaller values of $\bar\vartheta_v$, the strong CP bound on
$\mu$ is relaxed and the corresponding mass scale shifts to lower values,
illustrating the loss of the standard one-parameter prediction.

The mechanism discussed above modifies the relation between the axion mass
and the dark matter abundance while leaving the axion couplings
unchanged. In particular, the standard relation between $m_a$ and
$g_{a\gamma\gamma}$ remains intact; what changes is the region of
parameter space in which the QCD axion can account for the observed dark
matter.

Thus the post-inflationary scenario ceases to be fully predictive in terms
of $f_a$ alone unless the explicit breaking is negligible throughout the
relevant cosmological epoch.

We finally note that non-standard cosmological histories or early-time
modifications of QCD dynamics may alter the quantitative prediction for
the relic abundance~\cite{Dvali:1995ce,Visinelli:2009kt,Dvali:2026ceb}. In this work we
assume the standard thermal history.

\section{Conclusion}

In this work we have revisited the cosmological evolution of the QCD
axion in the presence of small explicit violations of the
Peccei--Quinn symmetry. Although such violations are strongly
constrained by the neutron electric dipole moment, they introduce an
additional mass scale $\mu$ which can become relevant in the early
Universe. This occurs because the QCD contribution to the axion mass
is strongly suppressed at high temperatures, so that even a small
temperature-independent breaking can dominate the early-time dynamics.

The explicit breaking introduces a new timescale of order $\mu^{-1}$
which can control the annihilation of the axion string network. When
the bias induced by the explicit breaking becomes comparable to the
restoring force associated with the string tension, the system
collapses before the QCD phase transition. In this regime the relic
abundance is no longer primarily controlled by QCD dynamics, and instead
scales parametrically as $\Omega_a \propto f_a\,\mu^{-1/2}$. As a
consequence, the usual one-to-one relation between the axion decay
constant and the dark matter abundance is lost.

Combining the strong CP constraint with the condition for early
annihilation, we find that this effect can operate for axion masses
$m_a \gtrsim 10^{-3}|\bar\vartheta_v|^{1/2}\,\mathrm{eV}$,
corresponding to
$f_a \lesssim 10^{10}|\bar\vartheta_v|^{-1/2}\,\mathrm{GeV}$,
where $\bar\vartheta_v$ denotes the initial CP-violating angle, i.e.\ the
value of the effective $\bar\vartheta$ parameter in the absence of the
axion.

The above bounds are sensitive to the initial CP-violating angle.
For $\bar\vartheta_v \sim \mathcal{O}(1)$, the strong CP constraint
enforces a strong suppression of the explicit breaking scale $\mu$,
leading to the parametric window identified above. For smaller values
of $\bar\vartheta_v$, the constraint on $\mu$ is relaxed and the allowed
range increases. Already for $\bar\vartheta_v \lesssim 10^{-4}$,
predictivity is effectively lost over the phenomenologically relevant
parameter space, while in the limit $\bar\vartheta_v \ll 1$ the strong CP
bound places only a weak restriction on $\mu$.

Importantly, this effect can arise in the region where post-inflationary
QCD axion dark matter is actively explored by haloscope experiments,
corresponding to axion masses in the $\mu\mathrm{eV}$ range and above.

Our results show that the usual one-parameter predictivity of
post-inflationary axion dark matter relies on the approximate validity of
the Peccei--Quinn symmetry over cosmological timescales. Even in the
minimal scenario the relic abundance carries uncertainties from string
network dynamics, but explicit PQ breaking introduces an additional and
independent source of model dependence. Very small violations, compatible
with the strong CP bound, can determine the cosmological evolution of the
axion and modify the standard relation between $f_a$ and the dark matter
abundance.
 In
this sense, axion dark matter can probe the ultraviolet quality of the
Peccei--Quinn symmetry, rather than being determined solely by infrared
QCD dynamics.

\,

\,

\noindent\textit{Acknowledgments.}
I thank Gia Dvali for useful comments regarding the evolution of
couplings in the early Universe, Goran Senjanović for ongoing
discussions on the strong CP problem and useful comments, and Luca Visinelli for reading the manuscript. I am further grateful to Marco Gorghetto for
useful comments on a preliminary version of this work. I also thank the anonymous Referee for comments that significantly improved
the presentation of this work.

\setlength{\bibsep}{4pt}
\bibliography{references}

@article{Blinov:2019jqc,
    author = "Blinov, Nikita and Dolan, Matthew J and Draper, Patrick",
    title = "{Imprints of the Early Universe on Axion Dark Matter Substructure}",
    eprint = "1911.07853",
    archivePrefix = "arXiv",
    primaryClass = "astro-ph.CO",
    reportNumber = "FERMILAB-PUB-19-560-A-T",
    doi = "10.1103/PhysRevD.101.035002",
    journal = "Phys. Rev. D",
    volume = "101",
    number = "3",
    pages = "035002",
    year = "2020"
}

@article{Arias:2023wyg,
    author = "Arias, Paola and Bernal, Nicol{\'a}s and Osi{\'n}ski, Jacek K. and Roszkowski, Leszek and Venegas, Moira",
    title = "{Revisiting signatures of thermal axions in nonstandard cosmologies}",
    eprint = "2308.01352",
    archivePrefix = "arXiv",
    primaryClass = "hep-ph",
    doi = "10.1103/PhysRevD.109.123529",
    journal = "Phys. Rev. D",
    volume = "109",
    number = "12",
    pages = "123529",
    year = "2024"
}

@article{Arias:2022qjt,
    author = "Arias, Paola and Bernal, Nicol{\'a}s and Osi{\'n}ski, Jacek K. and Roszkowski, Leszek",
    title = "{Dark matter axions in the early universe with a period of increasing temperature}",
    eprint = "2207.07677",
    archivePrefix = "arXiv",
    primaryClass = "hep-ph",
    reportNumber = "PI/UAN-2022-719FT",
    doi = "10.1088/1475-7516/2023/05/028",
    journal = "JCAP",
    volume = "05",
    pages = "028",
    year = "2023"
}

@article{Visinelli:2017imh,
    author = "Visinelli, Luca",
    title = "{Light axion-like dark matter must be present during inflation}",
    eprint = "1703.08798",
    archivePrefix = "arXiv",
    primaryClass = "astro-ph.CO",
    doi = "10.1103/PhysRevD.96.023013",
    journal = "Phys. Rev. D",
    volume = "96",
    number = "2",
    pages = "023013",
    year = "2017"
}

@article{Nelson:2018via,
    author = "Nelson, Ann E. and Xiao, Huangyu",
    title = "{Axion Cosmology with Early Matter Domination}",
    eprint = "1807.07176",
    archivePrefix = "arXiv",
    primaryClass = "astro-ph.CO",
    doi = "10.1103/PhysRevD.98.063516",
    journal = "Phys. Rev. D",
    volume = "98",
    number = "6",
    pages = "063516",
    year = "2018"
}

@article{Chang:2019tvx,
    author = "Chang, Chia-Feng and Cui, Yanou",
    title = "{New Perspectives on Axion Misalignment Mechanism}",
    eprint = "1911.11885",
    archivePrefix = "arXiv",
    primaryClass = "hep-ph",
    doi = "10.1103/PhysRevD.102.015003",
    journal = "Phys. Rev. D",
    volume = "102",
    number = "1",
    pages = "015003",
    year = "2020"
}

@article{Co:2018mho,
    author = "Co, Raymond T. and Gonzalez, Eric and Harigaya, Keisuke",
    title = "{Axion Misalignment Driven to the Hilltop}",
    eprint = "1812.11192",
    archivePrefix = "arXiv",
    primaryClass = "hep-ph",
    reportNumber = "LCTP-18-33",
    doi = "10.1007/JHEP05(2019)163",
    journal = "JHEP",
    volume = "05",
    pages = "163",
    year = "2019"
}

@article{Lyth:1991ub,
    author = "Lyth, D. H.",
    title = "{Axions and inflation: Sitting in the vacuum}",
    reportNumber = "LANC-TH-91-02-REV-2, LANC-TH-91-02",
    doi = "10.1103/PhysRevD.45.3394",
    journal = "Phys. Rev. D",
    volume = "45",
    pages = "3394--3404",
    year = "1992"
}

@article{Higaki:2016yqk,
    author = "Higaki, Tetsutaro and Jeong, Kwang Sik and Kitajima, Naoya and Takahashi, Fuminobu",
    title = "{Quality of the Peccei-Quinn symmetry in the Aligned QCD Axion and Cosmological Implications}",
    eprint = "1603.02090",
    archivePrefix = "arXiv",
    primaryClass = "hep-ph",
    reportNumber = "TU-1017, IPMU16-0030, APCTP-PRE-2016-006, PNUTP-16-A11",
    doi = "10.1007/JHEP06(2016)150",
    journal = "JHEP",
    volume = "06",
    pages = "150",
    year = "2016"
}

@article{Higaki:2015jag,
    author = "Higaki, Tetsutaro and Jeong, Kwang Sik and Kitajima, Naoya and Takahashi, Fuminobu",
    title = "{The QCD Axion from Aligned Axions and Diphoton Excess}",
    eprint = "1512.05295",
    archivePrefix = "arXiv",
    primaryClass = "hep-ph",
    reportNumber = "TU-1012, IPMU-15-0212, APCTP-PRE2015-029",
    doi = "10.1016/j.physletb.2016.01.055",
    journal = "Phys. Lett. B",
    volume = "755",
    pages = "13--16",
    year = "2016"
}

@article{Jeong:2022kdr,
    author = "Jeong, Kwang Sik and Matsukawa, Kohei and Nakagawa, Shota and Takahashi, Fuminobu",
    title = "{Cosmological effects of Peccei-Quinn symmetry breaking on QCD axion dark matter}",
    eprint = "2201.00681",
    archivePrefix = "arXiv",
    primaryClass = "hep-ph",
    reportNumber = "PNUTP-22-A11, TU-1143",
    doi = "10.1088/1475-7516/2022/03/026",
    journal = "JCAP",
    volume = "03",
    number = "03",
    pages = "026",
    year = "2022"
}

@article{Karananas:2025uhy,
    author = "Karananas, Georgios K. and Shaposhnikov, Mikhail and Zell, Sebastian",
    title = "{A non-compact QCD axion}",
    eprint = "2512.20290",
    archivePrefix = "arXiv",
    primaryClass = "hep-ph",
    month = "12",
    year = "2025"
}

@article{Dvali:2026ceb,
    author = "Dvali, Gia and Fitz, Sophia and Komisel, Lucy",
    title = "{Removing the Cosmological Bound on the Axion Scale via Confinement During Inflation}",
    eprint = "2603.28620",
    archivePrefix = "arXiv",
    primaryClass = "hep-ph",
    month = "3",
    year = "2026"
}

@article{Callan:1976je,
    author = "Callan, Jr., Curtis G. and Dashen, R. F. and Gross, David J.",
    editor = "Taylor, J. C.",
    title = "{The Structure of the Gauge Theory Vacuum}",
    reportNumber = "COO-2220-75",
    doi = "10.1016/0370-2693(76)90277-X",
    journal = "Phys. Lett. B",
    volume = "63",
    pages = "334--340",
    year = "1976"
}

@article{Peccei:1977hh,
    author = "Peccei, R. D. and Quinn, Helen R.",
    title = "{CP Conservation in the Presence of Instantons}",
    reportNumber = "ITP-568-STANFORD",
    doi = "10.1103/PhysRevLett.38.1440",
    journal = "Phys. Rev. Lett.",
    volume = "38",
    pages = "1440--1443",
    year = "1977"
}

@article{Weinberg:1977ma,
    author = "Weinberg, Steven",
    title = "{A New Light Boson?}",
    reportNumber = "HUTP-77/A074",
    doi = "10.1103/PhysRevLett.40.223",
    journal = "Phys. Rev. Lett.",
    volume = "40",
    pages = "223--226",
    year = "1978"
}

@article{Wilczek:1977pj,
    author = "Wilczek, Frank",
    title = "{Problem of Strong  $P$  and  $T$  Invariance in the Presence of Instantons}",
    reportNumber = "Print-77-0939 (COLUMBIA)",
    doi = "10.1103/PhysRevLett.40.279",
    journal = "Phys. Rev. Lett.",
    volume = "40",
    pages = "279--282",
    year = "1978"
}

@article{Preskill:1982cy,
    author = "Preskill, John and Wise, Mark B. and Wilczek, Frank",
    editor = "Srednicki, M. A.",
    title = "{Cosmology of the Invisible Axion}",
    reportNumber = "HUTP-82-A048, NSF-ITP-82-103",
    doi = "10.1016/0370-2693(83)90637-8",
    journal = "Phys. Lett. B",
    volume = "120",
    pages = "127--132",
    year = "1983"
}

@article{Abbott:1982af,
    author = "Abbott, L. F. and Sikivie, P.",
    editor = "Srednicki, M. A.",
    title = "{A Cosmological Bound on the Invisible Axion}",
    reportNumber = "PRINT-82-0695 (BRANDEIS)",
    doi = "10.1016/0370-2693(83)90638-X",
    journal = "Phys. Lett. B",
    volume = "120",
    pages = "133--136",
    year = "1983"
}

@article{Dine:1982ah,
    author = "Dine, Michael and Fischler, Willy",
    editor = "Srednicki, M. A.",
    title = "{The Not So Harmless Axion}",
    reportNumber = "UPR-0201T",
    doi = "10.1016/0370-2693(83)90639-1",
    journal = "Phys. Lett. B",
    volume = "120",
    pages = "137--141",
    year = "1983"
}

@article{Kibble:1976sj,
    author = "Kibble, T. W. B.",
    title = "{Topology of Cosmic Domains and Strings}",
    reportNumber = "ICTP/75/5",
    doi = "10.1088/0305-4470/9/8/029",
    journal = "J. Phys. A",
    volume = "9",
    pages = "1387--1398",
    year = "1976"
}

@article{Vilenkin:1982ks,
    author = "Vilenkin, A. and Everett, A. E.",
    title = "{Cosmic Strings and Domain Walls in Models with Goldstone and PseudoGoldstone Bosons}",
    doi = "10.1103/PhysRevLett.48.1867",
    journal = "Phys. Rev. Lett.",
    volume = "48",
    pages = "1867--1870",
    year = "1982"
}

@article{Lyth:1992tw,
    author = "Lyth, David H. and Stewart, Ewan D.",
    title = "{Axions and inflation: String formation during inflation}",
    reportNumber = "LANC-TH-92-03",
    doi = "10.1103/PhysRevD.46.532",
    journal = "Phys. Rev. D",
    volume = "46",
    pages = "532--538",
    year = "1992"
}

@article{Chang:1998tb,
    author = "Chang, Sanghyeon and Hagmann, C. and Sikivie, P.",
    title = "{Studies of the motion and decay of axion walls bounded by strings}",
    eprint = "hep-ph/9807374",
    archivePrefix = "arXiv",
    reportNumber = "UFIFT-HEP-98-12",
    doi = "10.1103/PhysRevD.59.023505",
    journal = "Phys. Rev. D",
    volume = "59",
    pages = "023505",
    year = "1999"
}

@article{Hiramatsu:2010yn,
    author = "Hiramatsu, Takashi and Kawasaki, Masahiro and Saikawa, Ken'ichi",
    title = "{Evolution of String-Wall Networks and Axionic Domain Wall Problem}",
    eprint = "1012.4558",
    archivePrefix = "arXiv",
    primaryClass = "astro-ph.CO",
    reportNumber = "ICRR-REPORT-577-2010-10, IPMU10-0221, YITP-10-110",
    doi = "10.1088/1475-7516/2011/08/030",
    journal = "JCAP",
    volume = "08",
    pages = "030",
    year = "2011"
}

@article{Hiramatsu:2012gg,
    author = "Hiramatsu, Takashi and Kawasaki, Masahiro and Saikawa, Ken'ichi and Sekiguchi, Toyokazu",
    title = "{Production of dark matter axions from collapse of string-wall systems}",
    eprint = "1202.5851",
    archivePrefix = "arXiv",
    primaryClass = "hep-ph",
    reportNumber = "ICRR-REPORT-608-2011-25, IPMU12-0025, YITP-12-9",
    doi = "10.1103/PhysRevD.85.105020",
    journal = "Phys. Rev. D",
    volume = "85",
    pages = "105020",
    year = "2012",
    note = "[Erratum: Phys.Rev.D 86, 089902 (2012)]"
}

@article{Kawasaki:2018bzv,
    author = "Kawasaki, Masahiro and Sekiguchi, Toyokazu and Yamaguchi, Masahide and Yokoyama, Jun'ichi",
    title = "{Long-term dynamics of cosmological axion strings}",
    eprint = "1806.05566",
    archivePrefix = "arXiv",
    primaryClass = "hep-ph",
    reportNumber = "RESCEU-8/18, RESCEU-8-18",
    doi = "10.1093/ptep/pty098",
    journal = "PTEP",
    volume = "2018",
    number = "9",
    pages = "091E01",
    year = "2018"
}

@article{Saikawa:2012uk,
    author = "Saikawa, Ken'ichi and Yamaguchi, Masahide",
    title = "{Evolution and thermalization of dark matter axions in the condensed regime}",
    eprint = "1210.7080",
    archivePrefix = "arXiv",
    primaryClass = "hep-ph",
    reportNumber = "ICRR-REPORT-632-2012-21",
    doi = "10.1103/PhysRevD.87.085010",
    journal = "Phys. Rev. D",
    volume = "87",
    number = "8",
    pages = "085010",
    year = "2013"
}

@article{Vaquero:2018tib,
    author = "Vaquero, Alejandro and Redondo, Javier and Stadler, Julia",
    title = "{Early seeds of axion miniclusters}",
    eprint = "1809.09241",
    archivePrefix = "arXiv",
    primaryClass = "astro-ph.CO",
    doi = "10.1088/1475-7516/2019/04/012",
    journal = "JCAP",
    volume = "04",
    pages = "012",
    year = "2019"
}

@article{DiLuzio:2020wdo,
    author = "Di Luzio, Luca and Giannotti, Maurizio and Nardi, Enrico and Visinelli, Luca",
    title = "{The landscape of QCD axion models}",
    eprint = "2003.01100",
    archivePrefix = "arXiv",
    primaryClass = "hep-ph",
    reportNumber = "DESY 20-036, DESY-20-036",
    doi = "10.1016/j.physrep.2020.06.002",
    journal = "Phys. Rept.",
    volume = "870",
    pages = "1--117",
    year = "2020"
}

@article{Turner:1985si,
    author = "Turner, Michael S.",
    title = "{Cosmic and Local Mass Density of Invisible Axions}",
    reportNumber = "FERMILAB-PUB-85-149-A, EFI-85-67-CHICAGO, FERMILAB-PUB-85-093-A",
    doi = "10.1103/PhysRevD.33.889",
    journal = "Phys. Rev. D",
    volume = "33",
    pages = "889--896",
    year = "1986"
}

@article{Visinelli:2009kt,
    author = "Visinelli, Luca and Gondolo, Paolo",
    title = "{Axion cold dark matter in non-standard cosmologies}",
    eprint = "0912.0015",
    archivePrefix = "arXiv",
    primaryClass = "astro-ph.CO",
    doi = "10.1103/PhysRevD.81.063508",
    journal = "Phys. Rev. D",
    volume = "81",
    pages = "063508",
    year = "2010"
}

@article{Hiramatsu:2012sc,
    author = "Hiramatsu, Takashi and Kawasaki, Masahiro and Saikawa, Ken'ichi and Sekiguchi, Toyokazu",
    title = "{Axion cosmology with long-lived domain walls}",
    eprint = "1207.3166",
    archivePrefix = "arXiv",
    primaryClass = "hep-ph",
    reportNumber = "ICRR-REPORT-620-2012-9, IPMU12-0140, YITP-12-58",
    doi = "10.1088/1475-7516/2013/01/001",
    journal = "JCAP",
    volume = "01",
    pages = "001",
    year = "2013"
}

@article{Buschmann:2021sdq,
    author = "Buschmann, Malte and Foster, Joshua W. and Hook, Anson and Peterson, Adam and Willcox, Don E. and Zhang, Weiqun and Safdi, Benjamin R.",
    title = "{Dark matter from axion strings with adaptive mesh refinement}",
    eprint = "2108.05368",
    archivePrefix = "arXiv",
    primaryClass = "hep-ph",
    doi = "10.1038/s41467-022-28669-y",
    journal = "Nature Commun.",
    volume = "13",
    number = "1",
    pages = "1049",
    year = "2022"
}

@article{Gorghetto:2020qws,
    author = "Gorghetto, Marco and Hardy, Edward and Villadoro, Giovanni",
    title = "{More axions from strings}",
    eprint = "2007.04990",
    archivePrefix = "arXiv",
    primaryClass = "hep-ph",
    doi = "10.21468/SciPostPhys.10.2.050",
    journal = "SciPost Phys.",
    volume = "10",
    number = "2",
    pages = "050",
    year = "2021"
}

@article{Gorghetto:2018myk,
    author = "Gorghetto, Marco and Hardy, Edward and Villadoro, Giovanni",
    title = "{Axions from Strings: the Attractive Solution}",
    eprint = "1806.04677",
    archivePrefix = "arXiv",
    primaryClass = "hep-ph",
    doi = "10.1007/JHEP07(2018)151",
    journal = "JHEP",
    volume = "07",
    pages = "151",
    year = "2018"
}

@article{Benabou:2024msj,
    author = "Benabou, Joshua N. and Buschmann, Malte and Foster, Joshua W. and Safdi, Benjamin R.",
    title = "{Axion Mass Prediction from Adaptive Mesh Refinement Cosmological Lattice Simulations}",
    eprint = "2412.08699",
    archivePrefix = "arXiv",
    primaryClass = "hep-ph",
    reportNumber = "FERMILAB-PUB-24-0912-T",
    doi = "10.1103/6v21-d6sj",
    journal = "Phys. Rev. Lett.",
    volume = "134",
    number = "24",
    pages = "241003",
    year = "2025"
}

@article{Dvali:2022fdv,
    author = "Dvali, Gia",
    title = "{Strong-$CP$ with and without gravity}",
    eprint = "2209.14219",
    archivePrefix = "arXiv",
    primaryClass = "hep-ph",
    month = "9",
    year = "2022"
}

@article{Shabalin:1979gh,
    author = "Shabalin, E. P.",
    title = "{THE ELECTRIC DIPOLE MOMENTS OF BARYONS IN THE KOBAYASHI-MASKAWA CP NONINVARIANT THEORY}",
    reportNumber = "ITEP-131-1979",
    journal = "Sov. J. Nucl. Phys.",
    volume = "32",
    pages = "228",
    year = "1980"
}

@article{Ellis:1978hq,
    author = "Ellis, John R. and Gaillard, Mary K.",
    title = "{Strong and Weak CP Violation}",
    reportNumber = "FERMILAB-PUB-78-066-T",
    doi = "10.1016/0550-3213(79)90297-9",
    journal = "Nucl. Phys. B",
    volume = "150",
    pages = "141--162",
    year = "1979"
}

@article{Ellis:1976fn,
    author = "Ellis, John R. and Gaillard, Mary K. and Nanopoulos, Dimitri V.",
    title = "{Lefthanded Currents and CP Violation}",
    reportNumber = "CERN-TH-2116",
    doi = "10.1016/0550-3213(76)90203-0",
    journal = "Nucl. Phys. B",
    volume = "109",
    pages = "213--243",
    year = "1976"
}

@article{Baker:2006ts,
    author = "Baker, C. A. and others",
    title = "{An Improved experimental limit on the electric dipole moment of the neutron}",
    eprint = "hep-ex/0602020",
    archivePrefix = "arXiv",
    doi = "10.1103/PhysRevLett.97.131801",
    journal = "Phys. Rev. Lett.",
    volume = "97",
    pages = "131801",
    year = "2006"
}

@article{Crewther:1979pi,
    author = "Crewther, R. J. and Di Vecchia, P. and Veneziano, G. and Witten, Edward",
    title = "{Chiral Estimate of the Electric Dipole Moment of the Neutron in Quantum Chromodynamics}",
    reportNumber = "CERN-TH-2735",
    doi = "10.1016/0370-2693(79)90128-X",
    journal = "Phys. Lett. B",
    volume = "88",
    pages = "123",
    year = "1979",
    note = "[Erratum: Phys.Lett.B 91, 487 (1980)]"
}

@article{Baluni:1978rf,
    author = "Baluni, Varouzhan",
    title = "{CP Violating Effects in QCD}",
    reportNumber = "MIT-CTP-726",
    doi = "10.1103/PhysRevD.19.2227",
    journal = "Phys. Rev. D",
    volume = "19",
    pages = "2227--2230",
    year = "1979"
}

@article{Jackiw:1976pf,
    author = "Jackiw, R. and Rebbi, C.",
    editor = "Taylor, J. C.",
    title = "{Vacuum Periodicity in a Yang-Mills Quantum Theory}",
    reportNumber = "MIT-CTP-548",
    doi = "10.1103/PhysRevLett.37.172",
    journal = "Phys. Rev. Lett.",
    volume = "37",
    pages = "172--175",
    year = "1976"
}

@article{diCortona:2016AxionPrecisely,
  author        = {Grilli di Cortona, Giovanni and Hardy, Edward and Pardo Vega, Javier and Villadoro, Giovanni},
  title         = {The QCD axion, precisely},
  journal       = {JHEP},
  volume        = {01},
  pages         = {034},
  year          = {2016},
  doi           = {10.1007/JHEP01(2016)034},
  eprint        = {1511.02867},
  archivePrefix = {arXiv},
  primaryClass  = {hep-ph},
  url           = {https://arxiv.org/abs/1511.02867}
}

@article{Gorghetto:2018Attractive,
  author        = {Gorghetto, Marco and Hardy, Edward and Villadoro, Giovanni},
  title         = {Axions from strings: the attractive solution},
  journal       = {JHEP},
  volume        = {07},
  pages         = {151},
  year          = {2018},
  doi           = {10.1007/JHEP07(2018)151},
  eprint        = {1806.04677},
  archivePrefix = {arXiv},
  primaryClass  = {hep-ph},
  url           = {https://arxiv.org/abs/1806.04677}
}

@article{Gorghetto:2019TopSus,
  author        = {Gorghetto, Marco and Villadoro, Giovanni},
  title         = {Topological susceptibility and QCD axion mass: QED and NNLO corrections},
  journal       = {JHEP},
  volume        = {03},
  pages         = {033},
  year          = {2019},
  doi           = {10.1007/JHEP03(2019)033},
  eprint        = {1812.01008},
  archivePrefix = {arXiv},
  primaryClass  = {hep-ph},
  url           = {https://arxiv.org/abs/1812.01008}
}

@article{Gorghetto:2021MoreAxions,
  author        = {Gorghetto, Marco and Hardy, Edward and Villadoro, Giovanni},
  title         = {More axions from strings},
  journal       = {SciPost Phys.},
  volume        = {10},
  number        = {2},
  pages         = {050},
  year          = {2021},
  doi           = {10.21468/SciPostPhys.10.2.050},
  eprint        = {2007.04990},
  archivePrefix = {arXiv},
  primaryClass  = {hep-ph},
  url           = {https://scipost.org/SciPostPhys.10.2.050}
}

@article{Witten:1979vv,
    author = "Witten, Edward",
    title = "{Current Algebra Theorems for the U(1) Goldstone Boson}",
    reportNumber = "HUTP-79/A014",
    doi = "10.1016/0550-3213(79)90031-2",
    journal = "Nucl. Phys. B",
    volume = "156",
    pages = "269--283",
    year = "1979"
}

@article{Veneziano:1979ec,
    author = "Veneziano, G.",
    title = "{U(1) Without Instantons}",
    reportNumber = "CERN-TH-2651",
    doi = "10.1016/0550-3213(79)90332-8",
    journal = "Nucl. Phys. B",
    volume = "159",
    pages = "213--224",
    year = "1979"
}

@article{Hindmarsh:2021zkt,
    author = "Hindmarsh, Mark and Lizarraga, Joanes and Lopez-Eiguren, Asier and Urrestilla, Jon",
    title = "{Comment on ''More Axions from Strings''}",
    eprint = "2109.09679",
    archivePrefix = "arXiv",
    primaryClass = "astro-ph.CO",
    month = "9",
    year = "2021"
}

@article{Dvali:2005an,
    author = "Dvali, Gia",
    title = "{Three-form gauging of axion symmetries and gravity}",
    eprint = "hep-th/0507215",
    archivePrefix = "arXiv",
    month = "7",
    year = "2005"
}

@article{Senjanovic:2020pnq,
    author = "Senjanovi{\'c}, Goran",
    title = "{Natural Philosophy versus Philosophy of Naturalness}",
    eprint = "2001.10988",
    archivePrefix = "arXiv",
    primaryClass = "hep-ph",
    doi = "10.1142/S0217732320300062",
    journal = "Mod. Phys. Lett. A",
    volume = "35",
    number = "18",
    pages = "2030006",
    year = "2020"
}

@article{Dvali:1995ce,
    author = "Dvali, G. R.",
    title = "{Removing the cosmological bound on the axion scale}",
    eprint = "hep-ph/9505253",
    archivePrefix = "arXiv",
    reportNumber = "IFUP-TH-21-95",
    month = "5",
    year = "1995"
}

\end{document}